\documentclass{ifacconf}

\usepackage{natbib} 
\usepackage{amsmath} 
\usepackage{amssymb}  
\usepackage{graphicx}
\usepackage{epstopdf}
\usepackage{amsmath}
\usepackage{mathtools}
\usepackage{dsfont}

\usepackage{enumerate}
\usepackage{tikz}
\usepackage{bm}
\usepackage{siunitx}
\DeclareMathOperator*{\argmin}{arg\,min}

\usepackage{balance}  
\usepackage{algorithm}
\usepackage{setspace}
\usepackage{caption}
\usepackage{varwidth}
\usepackage[noend]{algpseudocode}

\begin{document}
\begin{frontmatter}

\title{Task Decomposition for MPC: A Computationally Efficient Approach for Linear Time-Varying Systems\thanksref{footnoteinfo}} 

\thanks[footnoteinfo]{This research was sustained in part by fellowship support from the National Physical Science Consortium and the National Institute of Standards and Technology.}

\author[First]{Charlott Vallon} 
\author[First]{Francesco Borrelli} 

\address[First]{University of California at Berkeley, Berkeley, CA 94701, USA (e-mail: {charlott, fborrelli}@berkeley.edu).}

\begin{abstract}                
A Task Decomposition method for iterative learning Model Predictive Control (TDMPC) for linear time-varying systems is presented. 
We consider the availability of state-input trajectories which solve an original task $\mathcal{T}1$, and design a feasible MPC policy for a new task, $\mathcal{T}2$, using stored data from $\mathcal{T}1$. 
Our approach applies to tasks $\mathcal{T}2$ which are composed of subtasks contained in $\mathcal{T}1$.
In this paper we formally define the task decomposition problem, and provide a feasibility proof for the resulting policy.
The proposed algorithm reduces the computational burden for linear time-varying systems with piecewise convex constraints. 
Simulation results demonstrate the improved efficiency of the proposed method on a robotic path-planning task.
\end{abstract}


\end{frontmatter}

\section{Introduction}

Classical Iterative Learning Controllers (ILCs) aim to improve a system's closed-loop reference tracking performance at each iteration of a repeated task (\cite{AA, batchILC}).
Recent work has also explored reference-free ILC for applications whose goals are more appropriately defined via a performance metric, rather than a reference trajectory to track.
Examples include autonomous racing tasks (e.g. ``minimize lap time'') (\cite{rosolia2017autonomousrace, melanie}), or optimizing flight paths for tethered energy-harvesting systems (e.g. ``maximize average power generation'') (\cite{vermillion}).

In both classical and reference-free ILC, the controller uses data from previous iterations to improve future closed-loop performance with respect to the appropriate performance metric. At the very first iteration, these methods require either a reference trajectory to track or a feasible trajectory with which to initialize the control algorithm. 
If the task changes, a new trajectory must be designed, which can be difficult for complex tasks.

A variety of model-based methods have been suggested for finding feasible trajectories or policies for new tasks using stored data from related tasks.
One approach relies on building and adapting trajectory libraries. 
For example, \cite{nguyen2016dynamic} design a walking gait across stepping stones for a bipedal robot by linearly interpolating trajectories from a library of asymptotically stable periodic walking gaits.
The authors of \cite{Yang2019} consider a set of actions and corresponding motion primitives for iterative teleoperative tasks. Given a new user-provided input, probabilistic inferences are made over the respective set of locally feasible trajectories.
Similar approaches considering probabilistic distributions over trajectory libraries are proposed for robotic manipulators and autonomous vehicles in \cite{probabilisticprimitives} and \cite{zhi2019octnet}, respectively. 
In \cite{6}, environment features are used to divide a task and create a library of local trajectories in relative state space frames. These trajectories are then pieced back together according to the features of the new task environment. The authors in \cite{BG} propose running a desired path planning method in parallel with a retrieve and repair algorithm that adapts a reference trajectory from a previous task to the constraints of a new task.
While these methods decrease planning time, they require verifying or interpolating saved trajectories at each new time step, and cannot a priori guarantee constraint satisfaction.

Other approaches learn generalizable policies from stored task data. 
The authors of \cite{dai2018moments} consider linear hybrid systems. Data is collected in individual modes, and a polynomial optimization problem is formulated to find a stabilizing controller for arbitrary switching sequences. 
In \cite{pereida2018data}, a fixed map is learned between a given reference trajectory and the input sequence required to track the trajectory with a linear system. This defines a policy given a new reference trajectory, but does not provide the trajectory itself.
The authors of \cite{human-guided} learn a mapping between a robot gripper pose performing the same task with different tools, based on online human corrections. 
This method was effective in demonstrations, but required human supervision and cannot guarantee safety.

In this paper, our objective is to efficiently find a feasible trajectory to smartly initialize an Iterative Learning Model Predictive Controller (ILMPC) (\cite{Rosolia_IFAC17_LMPCLinear}) for a new task. 
ILMPC is a reference-free ILC that uses a \emph{safe set} to design an MPC policy for an iterative control task. This safe set is initialized using a feasible task trajectory, and collects states from which the task can be completed.
In \cite{mbttl}, a Task Decomposition for ILMPC (TDMPC) algorithm was introduced for nonlinear, constrained dynamical systems. 
TDMPC is data-efficient, requires no human supervision, and, if the algorithm converges, produces trajectories that are guaranteed to satisfy all constraints for the new task.
TDMPC decomposes an initial task $\mathcal{T}1$ into different modes of operation, called \emph{subtasks}, and adapts stored $\mathcal{T}1$ trajectories to a new task $\mathcal{T}2$ only at points of subtask transition, by solving one-step controllability problems.
The main contributions of this paper are as follows:
\begin{enumerate}[1.]
    \item We present an extension to the TDMPC algorithm in \cite{mbttl}. We introduce a new formulation for linear time-varying systems with piecewise convex state and input constraints. The new formulation further reduces the computational burden of finding feasible trajectories for a new task $\mathcal{T}2$ by formulating the goal as a convex optimization problem, and simultaneously increases the size of the resulting $\mathcal{T}2$ safe set.
    \item We prove that the induced safe set based MPC policy is feasible for $\mathcal{T}2$. This policy can be used to initialize an iterative learning control algorithm, or to directly obtain a suboptimal execution of $\mathcal{T}2$.
\end{enumerate}

\section{Problem Definition} \label{sec:PD} 
\subsection{Tasks and Subtasks}\label{ssec:ssilmpc}
Consider a discrete-time system with linear, time-varying dynamics
\begin{equation}\label{eq:gensystem}
    x_{k+1}=A_kx_k + B_ku_k,
\end{equation}
subject to state and input constraints 
\begin{equation}\label{eq:genconstraints}
    x_k \in \mathcal{X},~ u_k \in \mathcal{U},
\end{equation}
where the vectors $x_k$ and $u_k$ collect the states and inputs at time step $k$.
We define a set $\mathcal{P} \subset \mathcal{X}$ to be the target set for an iterative task $\mathcal{T}$, performed repeatedly by the system and defined by the tuple
\begin{align*}
    \mathcal{T} = \left\{ \mathcal{X}, \mathcal{U}, \mathcal{P}\right\}.
\end{align*}
In this work we consider tasks $\mathcal{T}$ that can be decomposed into an ordered sequence of subtasks with piecewise linear dynamics and convex constraint sets. Specifically, the $i$-th subtask $\mathcal{S}_i$ is the tuple
\begin{equation}\label{eq:subtaskdef}
    \mathcal{S}_i = \left\{A_i, B_i, \mathcal{X}_i, \mathcal{U}_i, \mathcal{R}_i\right\}.
\end{equation}
Within $\mathcal{S}_i$, the system is subject to linear dynamics 
\begin{equation}\label{eq:subtaskdynamics}
    x_{k+1} = A_ix_k + B_iu_k,
\end{equation}
and convex state and input constraints
\begin{align}\label{eq:subtaskconstraints}
    x_k \in \mathcal{X}_i,~ u_k \in \mathcal{U}_i,
\end{align}
where $\mathcal{X}_i \subseteq \mathcal{X}$ and $\mathcal{U}_i \subseteq \mathcal{U}$ are convex sets.
$\mathcal{R}_i$ is the set of transition states from $\mathcal{S}_i$ into the next subtask $\mathcal{S}_{i+1}$:
\begin{equation}\label{eq:transitionset}
    \mathcal{R}_i \subseteq \mathcal{X}_i = \{x \in \mathcal{X}_i : \exists u \in \mathcal{U}_i, ~ A_ix + B_iu \in \mathcal{X}_{i+1}\}.
\end{equation}
A successful \textit{subtask execution E($\mathcal{S}_i$)} of a subtask $\mathcal{S}_i$ is a trajectory of states and inputs evolving according to (\ref{eq:subtaskdynamics}), respecting state and input constraints (\ref{eq:subtaskconstraints}), and ending in the subtask transition set (\ref{eq:transitionset}). 
We define the $j$-th successful execution of subtask $\mathcal{S}_i$ as
\begin{subequations}
\begin{align}\label{eq:subtaskexec}
E^j(\mathcal{S}_i) & = [{\bf{x}}_i^j, {\bf{u}}_i^j], \\
{\bf{x}}_i^j & = [x_{0}^j,x_{1}^j,~...,x_{T_i^j}^j], ~x_k \in \mathcal{X}_i ~~ \forall k \in [0, T^j_i], \nonumber\\
&~~~~~~~~~~~~~~~~~~~~~~~~~~{x^j_{T_i^j}} \in \mathcal{R}_i, \label{eq:subtasktrans}\\
    {\bf{u}}_i^j & = [u_0^j,u_1^j,~...,u_{T_i^j}^j], ~u_k \in \mathcal{U}_i ~~ \forall k \in [0, T^j_i], \label{eq:subtaskinput} 
\end{align}
\end{subequations}
where $x_k^j$ and $u_k^j$ denote the system state and the control input at time $k$ of subtask execution $j$. $T_i^j$ is the duration of the $j$-th execution of subtask $i$.
Fig.~\ref{fig:subtasksetsformation} depicts three feasible subtask executions from a robotic path-planning example detailed in Sec.~\ref{sec:results}.
Again, the final state of each successful subtask execution is in the subtask transition set $\mathcal{R}_i$, from which it can evolve into the subsequent subtask. In order to keep notation  simple, we have written all subtask executions as beginning at time step $k=0$. 

Let task $\mathcal{T}$ be an ordered sequence of $M$ subtasks, $\mathcal{T}~=~\{\mathcal{S}_i \}_{i=1}^M$. The $j$-th successful task execution is the concatenation of the corresponding subtask executions:
\begin{align}\label{eq:taskexecution}
    E^j(\mathcal{T}) & = [E^j(\mathcal{S}_1), E^j(\mathcal{S}_2), ..., E^j(\mathcal{S}_M)] = [{\bf{x}}^j, {\bf{u}}^j],\\
    {\bf{x}}^j & = [{\bf{x}}_1^j,{\bf{x}}_2^j,~...,{\bf{x}}_M^j], \nonumber \\
    {\bf{u}}^j & = [{\bf{u}}_1^j,{\bf{u}}_2^j,~...,{\bf{u}}_M^j], 
     \nonumber \\
    x^j_{\alpha}& \in \mathcal{R}_i, ~~~~~~~~~~~~~~~~~~ i\in [1,M-1],\nonumber \\
    A_i&x^j_{\alpha} + B_i u^j_{\alpha}  \in \mathcal{X}_{i+1}, ~~ i\in [1,M-1],\nonumber 
    \end{align}
where $\alpha = T^j_{[1\rightarrow i]}$ is the duration of the first $i$ subtasks during the $j$-th task iteration.
When the state reaches a subtask transition set, the system has completed subtask $\mathcal{S}_i$, and it transitions into the following subtask $\mathcal{S}_{i+1}$. The task is completed when the system reaches the last subtask's transition set, $\mathcal{P} = \mathcal{R}_M$, the task's target set.

\textit{Definition:} A set $\mathcal{P}\subset \mathcal{X}$ is a \emph{control invariant set} for the system (\ref{eq:gensystem}) subject to the constraints (\ref{eq:genconstraints}) if:
\begin{equation}\label{eq:controlinvariant}
    x_k \in \mathcal{P} \implies \exists u_k \in \mathcal{U}~ : ~A_kx_k + B_ku_k \in \mathcal{P},~ \forall k \geq 0.
\end{equation}

\textit{Assumption 1}:
$\mathcal{P}$ is a control invariant set (\ref{eq:controlinvariant}).

The optimal task $\mathcal{T}$ completion problem is given by:
    \begin{align}\label{eq:minTimeOC}
        V^\star_{0 \rightarrow T} (x_0) = & \min_{T, u_0, ..., u_{T-1}} \sum_{k=0}^T h(x_k, u_k) \\
        & ~~~~~~ \mathrm{s.t.}~~~~~ x_{k+1} = f(x_k, u_k), \nonumber \\ 
        & ~~~~~~~~~~~~~~~ x_k \in \mathcal{X}, u_k \in \mathcal{U} ~~ \forall k \geq 0,\nonumber \\
        & ~~~~~~~~~~~~~~~ x_{T} \in \mathcal{P}, \nonumber
    \end{align}
where $V^\star_{0 \rightarrow T} (x_0)$ is the optimal cost-to-go from the initial state $x_0$, and $h(x_k, u_k)$ is a chosen stage cost. 

\subsection{Safe Set Based ILMPC}
In \cite{Rosolia_IFAC17_LMPCLinear}, a data-driven formulation for approximating the optimal control task (\ref{eq:minTimeOC}) for a linear time-invariant system with convex constraints (\ref{eq:genconstraints}) is introduced. Here we propose a formulation for linear \textit{time-varying} systems with \textit{piecewise} convex constraints. 

\begin{figure}
    \centering
    \includegraphics[width=1\columnwidth]{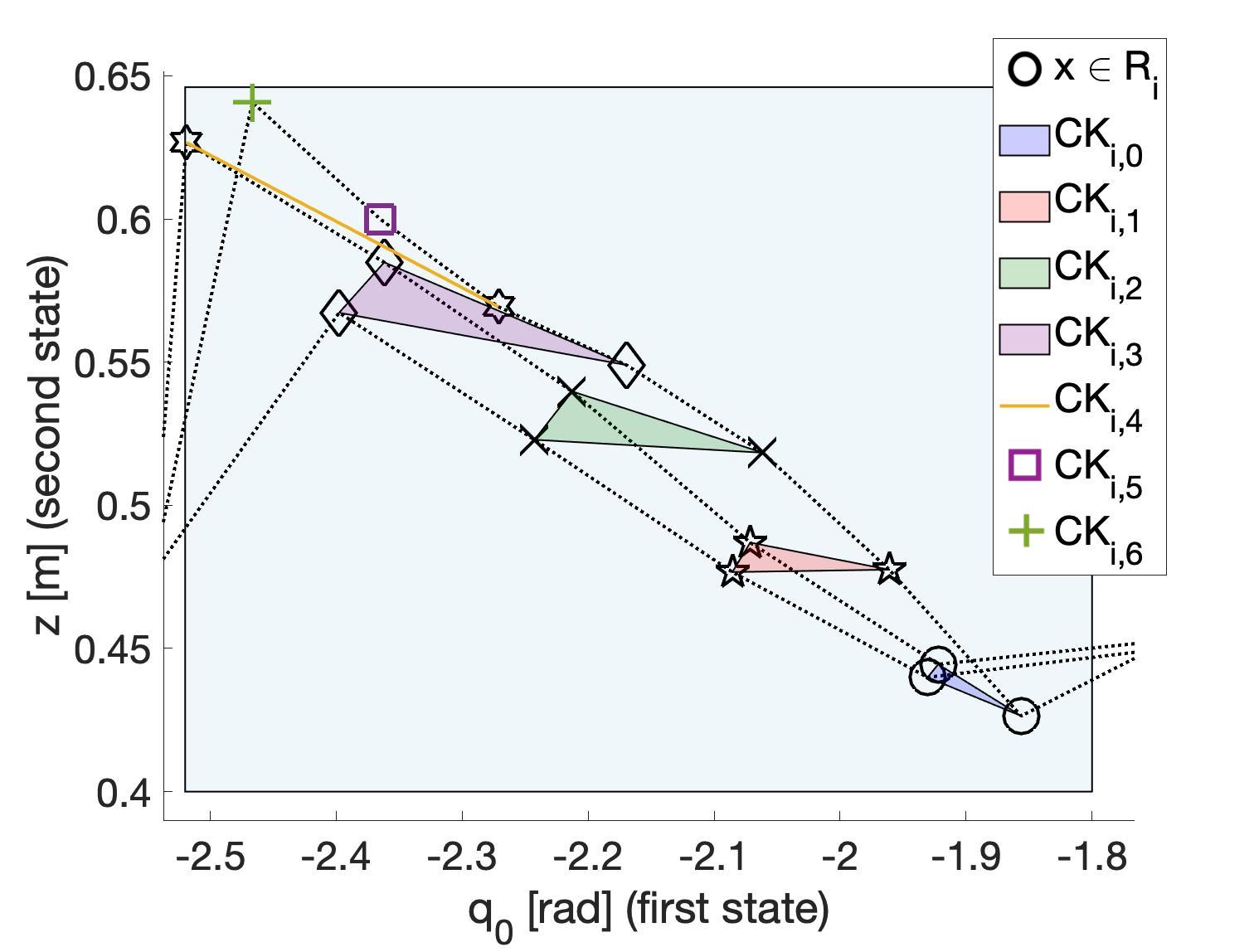}
    \caption{Convex subtask safe sets contain states from which the transition set can be reached in a certain number of steps.}
    \label{fig:subtasksetsformation}
\end{figure}

Each execution of task $\mathcal{T}$ is referred to as an iteration.
After $J$ number of task iterations, we define the time-indexed \emph{sampled subtask safe set} of subtask $\mathcal{S}_i$ as:
\begin{equation}\label{eq:SS}
\begin{aligned}
\mathcal{KS}_{i,k} = \textrm{}\left\{\bigcup_{j=1}^J x^j_{T^j_i - k} \right\},  ~ k \in [0, \max_j T^j_i -1]. 
\end{aligned}
\end{equation}
For given $k$ and $i$, $x^j_{T^j_i - k}$ is the $k$-to-last state visited in $\mathcal{S}_i$ during the $j$-th task iteration. Thus the set (\ref{eq:SS}) is the collection of states from which the system reaches the subtask transition set $\mathcal{R}_i$ in exactly $k$ steps during a previously recorded task iteration, while satisfying subtask constraints (\ref{eq:subtaskconstraints}).
We similarly define a time-indexed \emph{sampled subtask input set} of subtask $\mathcal{S}_i$ as:
\begin{equation*}
\begin{aligned}
\mathcal{KU}_{i,k} = \textrm{}\left\{\bigcup_{j=1}^J u^j_{T^j_i - k} \right\},  ~ k \in [0, \max_j T^j_i -1]. 
\end{aligned}
\end{equation*}
Because $\mathcal{X}_i$ and $\mathcal{U}_i$ are convex, there also exists a feasible $k$-step input sequence to $\mathcal{R}_i$ for each convex combination of elements in $\mathcal{KS}_{i,k}^J$. 
We define \emph{convex subtask safe sets} and \emph{convex subtask input sets} as: 
\begin{equation}\label{eq:convexss}
\begin{aligned}
    \mathcal{CK}_{i,k} & =  \left\{\sum_{p=1}^{|\mathcal{KS}_{i,k}|} \lambda_pz_p : \lambda_p \geq 0, \sum_{p=1}^{|\mathcal{KS}_{i,k}|} \lambda_p = 1, z_p \in \mathcal{KS}_{i,k} \right\}, \\
    \mathcal{CU}_{i,k} & =  \left\{\sum_{p=1}^{|\mathcal{KU}_{i,k}|} \lambda_pw_p : \lambda_p \geq 0, \sum_{p=1}^{|\mathcal{KU}_{i,k}|} \lambda_p = 1, w_p \in \mathcal{KU}_{i,k} \right\},
\end{aligned}
\end{equation}
where $|\mathcal{KS}_{i,k}|$ is the cardinality of $\mathcal{KS}_{i,k}$.
Fig.~\ref{fig:subtasksetsformation} depicts these sets for three trajectories ($J=3$) through a subtask from a robotic path-planning detailed in Sec.~\ref{sec:results}.
We define a barycentric cost-to-go over the convex subtask safe sets:
    \begin{align}\label{eq:barycost}
    v(x) &= \min_{\lambda_p \geq 0, ~I, ~K} ~ \sum_{p=0}^{|\mathcal{KS}_{I,K}|} \lambda_p V(z_p)  \\
    & ~~~~~~~~\mathrm{s.t. } ~  \sum_{p=0}^{|\mathcal{KS}_{I,K}|}  \lambda_p = 1,  \nonumber \\ 
    & ~~~~~~~~~~~~ \sum_{p=0}^{|\mathcal{KS}_{I,K}|} \lambda_p z_p = x, ~~ z_p \in \mathcal{KS}_{I,K}, \nonumber
    \end{align}
where $V(z_p)$ is the realized cost-to-go from state $z_p$ during a past execution.
At time step $k$ of iteration $J+1$, we approximate the optimal control problem (\ref{eq:minTimeOC}) by solving:
\begin{align}\label{eq:minTimeLMPC}
    &V(x_k^{J+1}) =  \\ 
    & \min_{u_{k|k}, ..., u_{k+N-1|k}, I, K} \sum_{t=k}^{k+N-1} h(x_{t|k}, u_{t|k}) + v(x_{k+N | k}) \nonumber \\
    & ~~~~~~~\mathrm{s.t.}~~~~~~~~~  x_{t+1 | k} = f(x_{t|k}, u_{t|k}, t), \forall t \in [k, k+N-1],\nonumber  \\
    & ~~~~~~~~~~~~~~~~~~~~ x_{t|k} \in \mathcal{X},~ u_{t|k} \in \mathcal{U},~~ \forall t \in [k, k+N-1], \nonumber \\
    & ~~~~~~~~~~~~~~~~~~~~ x_{k|k} = x_k^j,\nonumber \\ 
    & ~~~~~~~~~~~~~~~~~~~~ x_{k+N | k} \in \mathcal{CK}_{I,K} \cup \mathcal{P},\nonumber 
\end{align}
which searches for an input sequence over a horizon $N$ that controls the system (\ref{eq:gensystem}) to the state in a convex subtask safe state set or task target set $\mathcal{P}$ with the lowest cost-to-go (\ref{eq:barycost}). We use a receding horizon strategy:
\begin{align}\label{eq:ilmpc-policy}
    u(x^j_k) = \pi^{\mathrm{ILMPC}}(x^j_k)= u^\star_{k|k}.
\end{align}
At the first iteration of a new task, the ILMPC (\ref{eq:minTimeLMPC}) requires non-empty sets $\mathcal{CK}_{\cdot, \cdot}$ containing at least one feasible execution of the task. Next we present a computationally efficient approach for creating such sets using data from executions of a different, previous task. This new computationally efficient formulation of TDMPC for tasks with piecewise linear, convex modes is the main contribution of this work compared to \cite{mbttl}.
We will require the following definition of controllability. 

\textit{Definition:} For a given target set $\mathcal{R} \subset \mathcal{X}$, the \emph{N-step controllable set} $\mathcal{K}_N(\mathcal{R})$ of a system (\ref{eq:gensystem}) subject to constraints (\ref{eq:genconstraints}) is defined recursively as:
\begin{align*}
    \mathcal{K}_j(\mathcal{R}) & = \mathrm{Pre}(\mathcal{K}_{j-1}(\mathcal{R})) \cap \mathcal{X},~ \mathcal{K}_0(\mathcal{R}) = \mathcal{R},\\
    & ~~~~~~~~~~~~~~~~~~~~~~~~~~~~ j \in \left\{1, ..., N\right\}, \\
    \mathrm{Pre}(\mathcal{R}) &= \left\{x : \exists u \in \mathcal{U} : A_kx+B_ku \in \mathcal{R} \right\}.
\end{align*}
For all states in the $N$-step controllable set to $\mathcal{R}$ there exists a feasible input sequence such that the system will be driven into $\mathcal{R}$ in $N$ steps.
\begin{prop}\label{prop:1}
In general, not all states belonging to the convex hull of stored subtask executions are controllable to the subtask transition set (\ref{eq:transitionset}). 
\end{prop}
\begin{pf}
Proof in Appendix.
\end{pf}
\begin{prop}\label{prop:2}
All states in the time-indexed convex subtask safe sets (\ref{eq:convexss}) are controllable to the subtask transition set (\ref{eq:transitionset}).
\end{prop}
\begin{pf}
Proof follows from Thm.~\ref{th1} in Sec.~\ref{ssec:feasibility}.
\end{pf}

Propositions \ref{prop:1} and \ref{prop:2} motivate the approach proposed in this paper of storing task executions in time-indexed convex subtask safe sets (\ref{eq:convexss}).


\section{Task Decomposition for ILMPC}
Let Task~$1$  and Task~$2$ be different ordered sequences of the same $M$ subtasks:
\begin{equation}
\begin{aligned}
    \mathcal{T}1 &=\{ \mathcal{S}_i \}_{i=1}^M, ~~ \mathcal{T}2 =\{ \mathcal{S}_{l_i} \}_{i=1}^M, \label{eq:twotasks}
\end{aligned}
\end{equation}
where the sequence $[l_1, l_2, ...,l_M]$ is a reordering of the sequence $[1,2,...,M]$.
Assume non-empty subtask safe sets $\mathcal{KS}_{[1\rightarrow M]}$ (\ref{eq:SS}) containing task data from Task~$1$.

Our goal is to use stored subtask safe sets (\ref{eq:SS}) from Task~$1$ in order to find convex subtask safe sets (\ref{eq:convexss}) for Task~$2$.
These sets can then be used to initalize a controller for the new task.
The key intuition of TDMPC is that all successful subtask executions (\ref{eq:subtaskexec}) from Task~$1$ are also successful subtask executions for Task~$2$, as this definition only depends on properties (\ref{eq:subtasktrans}-\ref{eq:subtaskinput}) of each individual subtask, not the subtask sequence.
With this notion, Alg.~\ref{alg:MBTL} proceeds backwards through the new subtask sequence.

\subsection{TDMPC Algorithm}
The notation $(l_{M},\cdot)$ or $(i, \cdot)$ indicates that the described action is undertaken for all appropriate second arguments.
\begin{algorithm}
\caption{TDMPC algorithm}\label{alg:MBTL}
\begin{spacing}{1.2}
\begin{algorithmic}[1]

\State \begin{varwidth}[t]{\linewidth} \textbf{input} $\mathcal{KS}_{[1 \rightarrow M]}, ~\mathcal{KU}_{[1 \rightarrow M]}$, $[l_1, l_2, ..., l_M]$ \end{varwidth}

\State $\textbf{do} ~ \mathcal{CK}_{[1 \rightarrow M]}=\text{convexify}(\mathcal{KS}_{[1 \rightarrow M]}) ~~~~~~~~~~~~~~~~(\ref{eq:convexss})$

\State $ \textbf{do} ~\mathcal{SG}_i = \text{guard set clustering} (\mathcal{KS}_{[1 \rightarrow M]}) ~~~~~~~~~(\ref{eq:sampledguard})$

\State \textbf{initialize empty }$\hat{\mathcal{KS}}$, $\hat{\mathcal{KU}}$, $\hat{\mathcal{CK}}$, $\hat{\mathcal{CU}}$

\State $ \hat{\mathcal{CK}}_{l_M, \cdot} \gets~ {\mathcal{CK}}_{l_M, \cdot},~ \hat{\mathcal{CU}}_{l_M, \cdot} \gets~ {\mathcal{CU}}_{l_M, \cdot}$ 

\For{$i \in [l_{M-1} : -1 : l_1]$} 
\State $\hat{\mathcal{KS}}_{i, \cdot} \gets~ {\mathcal{KS}}_{i, \cdot}$ 
\State $\hat{\mathcal{KU}}_{i, \cdot} \gets~ {\mathcal{KU}}_{i, \cdot}$ 
\State \textbf{initialize empty } $\hat{\mathcal{KU}}_{i, 0}$

\For{$x \in \mathcal{SG}_i$}

\State $\textbf{check } (q^\star, u^\star) = \mathrm{Ctrb}(x, \hat{\mathcal{CK}}_{i+1, \cdot})~~~~~~~~~~~~~~~~~~(\ref{eq:controllability})$
\If {\textbf{infeasible}}

\State $\hat{\mathcal{KS}}_{i, \cdot} \gets \hat{\mathcal{KS}}_{i, \cdot} \text{\textbackslash} \mathrm{trajectory}(x)$
\State $\hat{\mathcal{KU}}_{i, \cdot} \gets \hat{\mathcal{KU}}_{i, \cdot} \text{\textbackslash} \mathrm{trajectory}(u^\star)$

\Else 
\State $\hat{\mathcal{KU}}_{i, 0} \gets u^\star $
\EndIf
\EndFor

\State $\hat{\mathcal{CK}}_{i, \cdot} \gets \text{convexify}(\hat{\mathcal{KS}}_{i, \cdot})~~~~~~~~~~~~~~~~~~~~~~~~~~(\ref{eq:convexss})$
\State $\hat{\mathcal{CU}}_{i, \cdot} \gets \text{convexify}(\hat{\mathcal{KU}}_{i, \cdot})~~~~~~~~~~~~~~~~~~~~~~~~~~(\ref{eq:convexss})$

\EndFor

\State \textbf{Return~ } $\mathcal{CK}_{[l_1 \rightarrow l_M]} \gets \hat{\mathcal{CK}},~ \mathcal{CU}_{[l_1 \rightarrow l_M]} \gets \hat{\mathcal{CU}}$

\end{algorithmic}
\end{spacing}
\end{algorithm}

\begin{itemize}
\item Consider the last subtask, $\mathcal{S}_{l_M}$. 
By definition, for any state in $\mathcal{CK}_{l_M, \cdot}$ there exists a stored input sequence in $\mathcal{CU}_{l_M, \cdot}$ that can be applied such that the system evolves into $\mathcal{R}_{l_M}$. (\underline{Algorithm \ref{alg:MBTL}, Lines 2-5}).

\item Prune the convex safe sets of the preceding subtask
$\mathcal{S}_{l_{M-1}}$ to contain only states which can also be controlled to $\mathcal{R}_{l_M}$. We verify this property only for states in the sampled guard set of $\mathcal{S}_{l_{M-1}}$, defined as:
\begin{equation}\label{eq:sampledguard}
    \mathcal{SG}_{l_{M-1}} = \mathcal{KS}_{l_{M-1},0}.
\end{equation}
The sampled guard set for subtask $l_{M-1}$ contains the states in $\mathcal{S}_{l_{M-1}}$ from which the system transitioned into another subtask during a past execution of $\mathcal{T} 1$ (\underline{Algorithm \ref{alg:MBTL}, Line 10}).

\item 
Determine which points in $\mathcal{SG}_{l_{M-1}}$ are one-step controllable to $\mathcal{CK}_{l_{M}, k}$ for some time index $k$. This problem can be solved using a variety of numerical approaches. In the results presented in this paper, we check controllability for each $k$, and choose the input $u$ to minimize the cost of the resulting state according to (\ref{eq:barycost}). Specifically, for each point $x \in \mathcal{SG}_{l_{M-1}}$ and index $k$, we solve $(q^\star, u^\star)  = \mathrm{Ctrb}(x, \hat{\mathcal{CK}}_{l_M, k})$, where: 
\begin{align}\label{eq:controllability}
    u^\star = & \argmin_{u} v(z)  \\
    & ~ \mathrm{s.t.} ~ z = A_{l_{M-1}}x + B_{l_{M-1}}u\nonumber  \\
    & ~~~~~~ u \in \mathcal{U}_{l_{M-1}} \nonumber \\
    & ~~~~~~ z \in \hat{\mathcal{CK}}_{l_M, k}, \nonumber \\
    q^\star = &~ v(A_{l_{M-1}}x + B_{l_{M-1}}u^\star).
\end{align}
If (\ref{eq:controllability}) is feasible, the previously stored $\mathcal{T}1$ cost-to-go (\ref{eq:barycost}) from the state $x$ is replaced by $q^\star$, the cost to reach the goal set of $\mathcal{T}2$.
(\underline{Algorithm \ref{alg:MBTL}, Line 11})

\item 
For all states $x$ in $\mathcal{SG}_{l_{M-1}}$ not controllable to any convex safe set in $\mathcal{S}_{l_{M}}$,
we remove the stored subtask execution ending in $x$ out of the set of subtask safe sets for $\mathcal{S}_{l_{M}}$.  (\underline{Algorithm \ref{alg:MBTL}, Lines 12-16})

\item 
After checking the entire sampled guard set, all remaining convex subtask safe sets for $\mathcal{S}_{l_{M-1}}$ are controllable to convex subtask safe sets in $\mathcal{S}_{l_M}$, and therefore also to $\mathcal{R}_{l_M}$.
(\underline{Algorithm \ref{alg:MBTL}, Lines 17-19})
\end{itemize}

Alg.~\ref{alg:MBTL} iterates backwards through the remaining subtasks, verifying the controllability of points in sampled guard sets to a convex subtask safe set in the following subtask.
The algorithm returns convex subtask safe sets for Task~$2$ that can be used to initialize an ILMPC (\ref{eq:minTimeLMPC} - \ref{eq:ilmpc-policy}) for Task~$2$.

TDMPC offers a computationally efficient, data-driven method of initializing an ILMPC for new tasks. 
In contrast to multi-step or set-based methods that are common for model-based or hybrid systems, Alg.~\ref{alg:MBTL} only solves controllability problems from discrete points in the sampled guard set to already verified feasible sets. 
Furthermore, TDMPC directly provides a robust policy for solving the task associated with the verified trajectories.

Note the reformulation of the stored Task~$1$ executions (\ref{eq:taskexecution}) into convex sets (\ref{eq:convexss}). We can thus replace the point-to-point controllability verification from \cite{mbttl} with point-to-set controllability in Alg.~\ref{alg:MBTL}. This allows for three major improvements to the procedure: 
\begin{enumerate}[1.]
    \item (\ref{eq:controllability}) is a convex optimization problem, which is, in general, much faster to solve than the non-convex point-to-point controllability.
    \item By using the convex hull of stored states (\ref{eq:convexss}) as a target set in (\ref{eq:controllability}), rather than individual states, more points in the sampled guard sets (\ref{eq:sampledguard}) can potentially be demonstrated to lead to feasible Task~$2$ executions.
    \item We increase the number of points for which we know a feasible Task~$2$ policy, since we implicitly consider all points in the time-indexed convex hulls of Task~$1$ trajectories (\ref{eq:convexss}), rather than only the Task~$1$ trajectories. 
\end{enumerate}

\subsection{Feasibility}\label{ssec:feasibility}
We prove the feasibility of ILMPC policies (\ref{eq:ilmpc-policy}) initialized using Alg.~\ref{alg:MBTL}. 

\emph{Assumption 2}:\label{ass:convextasks}
Task~$1$ and Task~$2$ are defined as in (\ref{eq:twotasks}), with each subtask defined by linear dynamics (\ref{eq:subtaskdynamics}) and convex constraints (\ref{eq:subtaskconstraints}).

\setcounter{thm}{0}
\begin{thm}\label{th1} Let Assumptions $1$-$2$ hold. Assume Alg.~\ref{alg:MBTL} outputs non-empty sets $\mathcal{CK}^0_{[l_1\rightarrow l_M]}$ for Task~$2$. Then, if $x_0 \in \mathcal{CK}^0_{[l_1\rightarrow l_M]}$, the policy $\pi^{\mathrm{ILMPC}}_{[l_1 \rightarrow l_M]}$, as defined in (\ref{eq:ilmpc-policy}), produces a feasible execution of Task~$2$.
\end{thm}

\begin{pf}
We will use induction to prove the feasibility.
First, we show that the ILMPC (\ref{eq:minTimeLMPC}-\ref{eq:ilmpc-policy}) is feasible at time step $k=0$ of the $j$-th execution of Task~$2$.
    By assumption $\mathcal{CK}^0_{[l_1\rightarrow l_M]}$ is not empty. From (\ref{eq:convexss}) we have that $\mathcal{CK}^0_{[l_1\rightarrow l_M]} \subseteq \mathcal{CK}^{j-1}_{[l_1\rightarrow l_M]} ~ \forall j \geq 1$, and consequently $\mathcal{CK}^{j-1}_{[l_1\rightarrow l_M]}$ is not empty. Therefore, $x_0 \in \mathcal{CK}^0_{[l_1\rightarrow l_M]} \subseteq \mathcal{CK}^{j-1}_{[l_1\rightarrow l_M]}$, and there exist $I^\star$, $K^\star$, and multipliers $\lambda_p^\star$ such that
    \begin{equation*}
        x_0 = \sum_{p=1}^{|\mathcal{CK}^{j-1}_{I^\star,K^\star}|} \lambda_p^\star x_p,~ x_p \in \mathcal{CK}^{j-1}_{I^\star,K^\star}.
    \end{equation*}
    We define 
    \begin{equation}\label{eq:ubar}
    \bar{u}= \lambda_p^\star u_p \in \mathcal{U}_{I^\star}
    \end{equation}
    where $u_p$ is the input associated with the state $x_p \in \mathcal{CK}^{j-1}_{I^\star,K^\star}$ in a previous task execution of Task~$1$.
    Now, note that we have
    \begin{equation}
        \begin{aligned}\label{eq:xbar}
        \bar{x} &= A_{I^\star} + B_{I^\star}\bar{u} = \sum_{p=1}^{|\mathcal{CK}^{j-1}_{I^\star,K^\star-1}|} \lambda_p^\star x_p, \\ 
        & x_p \in 
        \begin{cases}
        \mathcal{CK}^{j-1}_{I^\star,K^\star-1}, & K^\star \geq 1;\\
        \mathcal{CK}^{j-1}_{I^\star+1,K^{\star\star}} & K^\star = 0,
        \end{cases}
        \end{aligned}
    \end{equation}
    for some $K^{\star\star}$. The second case ($K^\star = 0$) follows directly from Alg.~ \ref{alg:MBTL}.
    This procedure (\ref{eq:ubar}-\ref{eq:xbar}) can be repeated $N$ times in order to find a feasible input sequence for the initial state $x_0$ satisfying (\ref{eq:minTimeLMPC}). Therefore there exists a feasible solution to the ILMPC (\ref{eq:minTimeLMPC} - \ref{eq:ilmpc-policy}) at time step $k=0$ of the $j$-th execution of Task~$2$.
    
    Next, we show that the policy is recursively feasible.
    Assume that at time step $k$ of the $j$-th iteration the ILMPC (\ref{eq:minTimeLMPC} - \ref{eq:ilmpc-policy}) is feasible, and let ${\bf{x}}_{k:k+N | k}^{\star, j}$ and ${\bf{u}}_{k:k+N | k}^{\star, j}$ be the optimal trajectory and input sequence according to (\ref{eq:minTimeLMPC}). From (\ref{eq:ilmpc-policy}), the realized state and input at time $k$ of the $j$-th iteration are given by
    \begin{equation*}
        \begin{aligned}
        x_k^j = x_{k|k}^{\star, j}, ~ u_k^j = u_{k|k}^{\star, j},
        \end{aligned}
    \end{equation*}
    and the terminal constraint in (\ref{eq:minTimeLMPC}) enforces $x_{k+N | k}^{\star, j} \in \mathcal{CK}^{j-1}_{I', K'}$ for some $I'$, $K'$, where $\mathcal{CK}^{j-1}_{I', K'}$ contains states from the previous $j-1$ trajectories. 
    As in (\ref{eq:ubar}-\ref{eq:xbar}), define an input and corresponding state
    \begin{equation*}
    \begin{aligned}
     u' &= \lambda_p^\star u_p \in \mathcal{U}_{I'} \\
     x' &= A_{I'}x_{k+N | k}^{\star, j} + B_{I'}u' \in \mathcal{CK}^{j-1}_{I',K'-1}. 
    \end{aligned}
    \end{equation*}
    We therefore have
    \begin{equation*}
        x^j_{k+1} = x_{k+1 | k}^{\star, j}.
    \end{equation*}
    It follows that at time step $k+1$ of the $j$-th Task~$2$ execution, the input sequence and related feasible state trajectory 
    \begin{equation}\label{eq:prooftrajectory}
        \begin{aligned}
        &[u_{k+1|k}^{\star, j},~u_{k+2|k}^{\star, j},~...,u_{k+N-1|k}^{\star, j}, u'] \\ 
        &[x_{k+1|k}^{\star, j},~x_{k+2|k}^{\star, j},~...,x_{k+N-1|k}^{\star, j}, x']
        \end{aligned}
    \end{equation}
    satisfy input and state constraints in (\ref{eq:minTimeLMPC}). Therefore, (\ref{eq:prooftrajectory}) is a feasible solution for (\ref{eq:minTimeLMPC}) at time step $k+1$.
    
    We have shown that at the $j$-th iteration of Task~$2$, $\forall j \geq 1$, the ILMPC is feasible at time step $k=0$ and that if the ILMPC is feasible at time step $k$, it must also be feasible at time step $k+1$. We can conclude by induction that (\ref{eq:minTimeLMPC} - \ref{eq:ilmpc-policy}) is feasible $\forall j\geq 1$ and $k \in \mathbb{Z}_{0+}$ when initialized with sets output by Alg.~\ref{alg:MBTL}.  \hfill$\blacksquare$
\end{pf}
It follows from the same arguments (\ref{eq:ubar}-\ref{eq:xbar}) that the ILMPC policy (\ref{eq:minTimeLMPC}-\ref{eq:ilmpc-policy}) will eventually bring any $x_0 \in \mathcal{CK}^J$ to the task target set $\mathcal{R}_{l_M}$.
	
\section{Results: Robot Path Planning}\label{sec:results}

\subsection{Task Formulation}
\begin{figure}
    \centering
    \includegraphics[width = 0.45\textwidth]{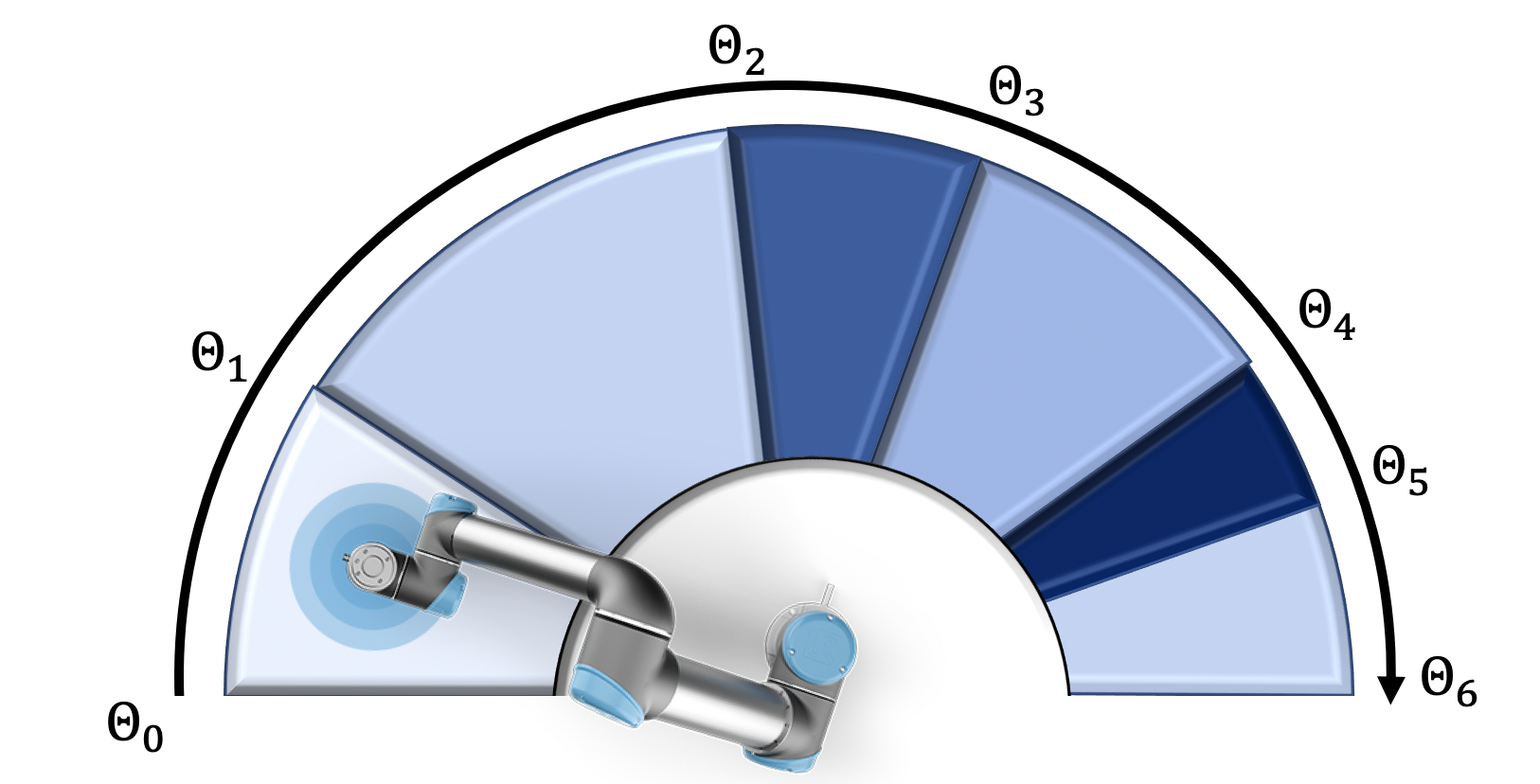}
    \caption{Topview of the path planning task. Each subtask corresponds to a pair of upper and lower obstacles.}
    \label{fig:robottaskpic}
\end{figure} 
We demonstrate the effectiveness of Alg.~\ref{alg:MBTL} in a robotic path planning example. 
Consider a UR5e\footnote{https://www.universal-robots.com/products/ur5-robot/} robotic arm tasked with maneuvering through six sets of obstacles modeled as extruded disks of varying heights above and below the robot. Each set of upper and lower obstacles leaves a workspace between the disks for the robot to move between. 
Here, each subtask $\mathcal{S}_i$ corresponds to the workspace between a pair of lower and upper obstacles.
Different subtask orderings correspond to a rearranging of the obstacle locations, indicated by $\Theta_i$ in Fig.~\ref{fig:robottaskpic}.

The UR5e has high end-effector reference tracking accuracy, allowing us to use a simplified end-effector model in place of a discretized second-order model as in \cite{robothard3}. We solve the task in the reduced state space:
\begin{subequations}
\begin{align}
    {x}_k & = [q_{0_k} ~ \dot{q}_{0_k} ~ z_k ~ \dot{z}_k]^\top, \nonumber \\
    {u}_k &= [\ddot{q}_{0_k} ~ \ddot{z}_k]^\top, \nonumber
\end{align}
\end{subequations}
where $q_{0_k}$ is the angle of the robot's base joint along the $\Theta$ direction and $z_k$ is the height of the robot end-effector at time step $k$, calculated from the six joint angles via forward kinematics. $\dot{q}_{0_k}$ and $\dot{z}_k$ are the corresponding velocities. We control $\ddot{q}_{0_k}$ and $\ddot{z}_k$, the accelerations of $q_0$ and $z$, respectively. This reduced state space allows us to formulate the task as a concatenation of $M=6$ subtasks with piecewise affine dynamics and convex constraints, according to (\ref{eq:subtaskdef}).

\subsubsection{Subtask Dynamics $A_i, B_i$}
We model the base-and-end-effector system as a quadruple integrator:
\begin{subequations}\label{eq:simplifiedRobotModel}
\begin{align}
    {x}_{k+1} & = A_i x_k + B_i u_k \\ 
    A_i &=
    \begin{bmatrix}1 & dt & 0 & 0 \\ 0 & 1 & 0 & 0\\ 0 & 0 & 1 & dt \\ 0 & 0 & 0 & 1 \end{bmatrix}, B_i = \begin{bmatrix}0 & 0 \\ dt & 0 \\ 0 & 0 \\ 0 & dt \end{bmatrix},~ \forall i \in [1,6]
\end{align}
\end{subequations}
where $dt = 0.01$ seconds is the sampling time. This simplified model holds as long as we operate within the region of high end-effector reference tracking accuracy, characterized in previous experiments. 

\subsubsection{Subtask Constraints $\mathcal{X}_i$}
\begin{equation}
\begin{aligned}
    {\mathcal{X}}_i =  \begin{bmatrix}\Theta_{i-1}~ \mathrm{rad} \\ -\pi ~ \mathrm{rad/s} \\ o_{\mathrm{min}, i} ~\mathrm{m} \\ \dot{z}_{\mathrm{min},i} ~ \mathrm{m/s} \end{bmatrix} \leq  \begin{bmatrix}q_0 \\ \dot{q}_{0k} \\ z_k \\ \dot{z}_k  \end{bmatrix}  \leq   \begin{bmatrix}\Theta_i ~ \mathrm{rad}\\ \pi ~ \mathrm{rad/s} \\ o_{\mathrm{max},i}~ \mathrm{m} \\ \dot{z}_{\mathrm{max},i} ~
    \mathrm{m/s} \end{bmatrix} \nonumber
\end{aligned}
\end{equation}
where $\Theta_{i-1}$ and $\Theta_{i}$ mark the cumulative angle to the beginning and end of the $i$-th obstacle, as in Fig.~\ref{fig:robottaskpic}.
The robot end-effector is constrained to remain in the space between the upper and lower obstacles, bounded by $o_{\mathrm{min}, i}$ and $o_{\mathrm{max}, i}$.
The base's rotational velocity $\dot{q}_{0k}$ and $\dot{z}_k$ are constrained to lie in the experimentally determined region of high end-effector tracking accuracy. Specifically, we take
\begin{subequations}
\begin{align*}
    \dot{z}_{\mathrm{max},i} & =  C_1\sin\bigg(\arccos\bigg(\frac{o_{\mathrm{min},i}  }{d_1}\bigg)\bigg), \\
    \dot{z}_{\mathrm{min},i} &= -\dot{z}_{\mathrm{max},i},
\end{align*}
\end{subequations}
where the constants $C_1$ and $d_1$ depend on setup parameters and joint limits provided by the manufacturer.

\subsubsection{Subtask Input Space $\mathcal{U}_i$}
\begin{equation}
 {\mathcal{U}}_i = \begin{bmatrix} -\pi ~ \mathrm{rad/s^2} \\ \ddot{z}_{\mathrm{min},i} ~ \mathrm{m/s^2} \end{bmatrix} \leq  \begin{bmatrix} \ddot{q}_{0k} \\ \ddot{z}_k  \end{bmatrix}  \leq   \begin{bmatrix}\pi ~ \mathrm{rad/s^2} \\ \ddot{z}_{\mathrm{max},i} ~ \mathrm{m/s^2} \end{bmatrix}, \nonumber
\end{equation}
where $\ddot{q}_{0k}$ and $\ddot{z}_k$ are constrained to lie in the experimentally determined region of high end-effector tracking accuracy. Specifically,
\begin{subequations}
\begin{align*}
    \ddot{z}_{\mathrm{max},i} & = C_2\sin\bigg(\arccos\bigg(\frac{o_{\mathrm{min},i}  }{d_2}\bigg) +  \frac{o_{\mathrm{min},i}  }{d_3}   \bigg), \\
    \ddot{z}_{\mathrm{min},i} &= -\ddot{z}_{\mathrm{max},i},
\end{align*}
\end{subequations}
where $C_2$, $d_2$ and $d_3$ depend on setup parameters and joint limits provided by the manufacturer.

\subsubsection{Subtask Transition Set, $\mathcal{R}_i$}
We define the subtask transition set to be the states along the subtask border where the next obstacle begins:
\begin{equation}
    \mathcal{R}_i = \{x \in \mathcal{X}_i: \exists u\in\mathcal{U}_i \text{, s.t. } q_0^+ \geq \Theta_{i}~ \}, \nonumber
\end{equation}
where $x^+ = A_ix+B_iu$ (\ref{eq:subtaskdynamics}).
The task target set is the end of the last mode:
\begin{equation}
    \mathcal{R}_6 = \left\{ x : q_0 = \Theta_6,~o_{\mathrm{min},6} \leq z \leq o_{\mathrm{max},6} \right \}.\nonumber
\end{equation}
The task goal is to reach the target set as quickly as possible:
\begin{equation}
    h(x_k, u_k) = \begin{cases}
			0, & x_k \in \mathcal{R}_6\\
            1, & \mathrm{otherwise}.
		 \end{cases}\nonumber
\end{equation}

\subsection{Experimental Results}
We evaluate the efficiency of Alg.~\ref{alg:MBTL} by comparing its run-time with the the run-time of the point-to-point controllability analysis for task decomposition introduced in \cite{mbttl} for nonlinear systems.

First, an ILMPC (\ref{eq:minTimeLMPC})-(\ref{eq:ilmpc-policy}) is used to complete five executions of five different training tasks, where each training task is a different reordering of the six obstacles. 
Each ILMPC is initialized with a suboptimal state and input trajectory that tracks the center-height of each subtask while the robot arm rotates at a low base velocity $\dot{q}_0$. 
In each task, the ILMPC tries to reach the target set as quickly as possible while avoiding the obstacles. 
The iterations are completed in simulation using the simplified model (\ref{eq:simplifiedRobotModel}), and the corresponding subtask executions are saved as convex subtask safe sets (\ref{eq:convexss}).
\begin{figure}
    \centering
    \includegraphics[width=\columnwidth]{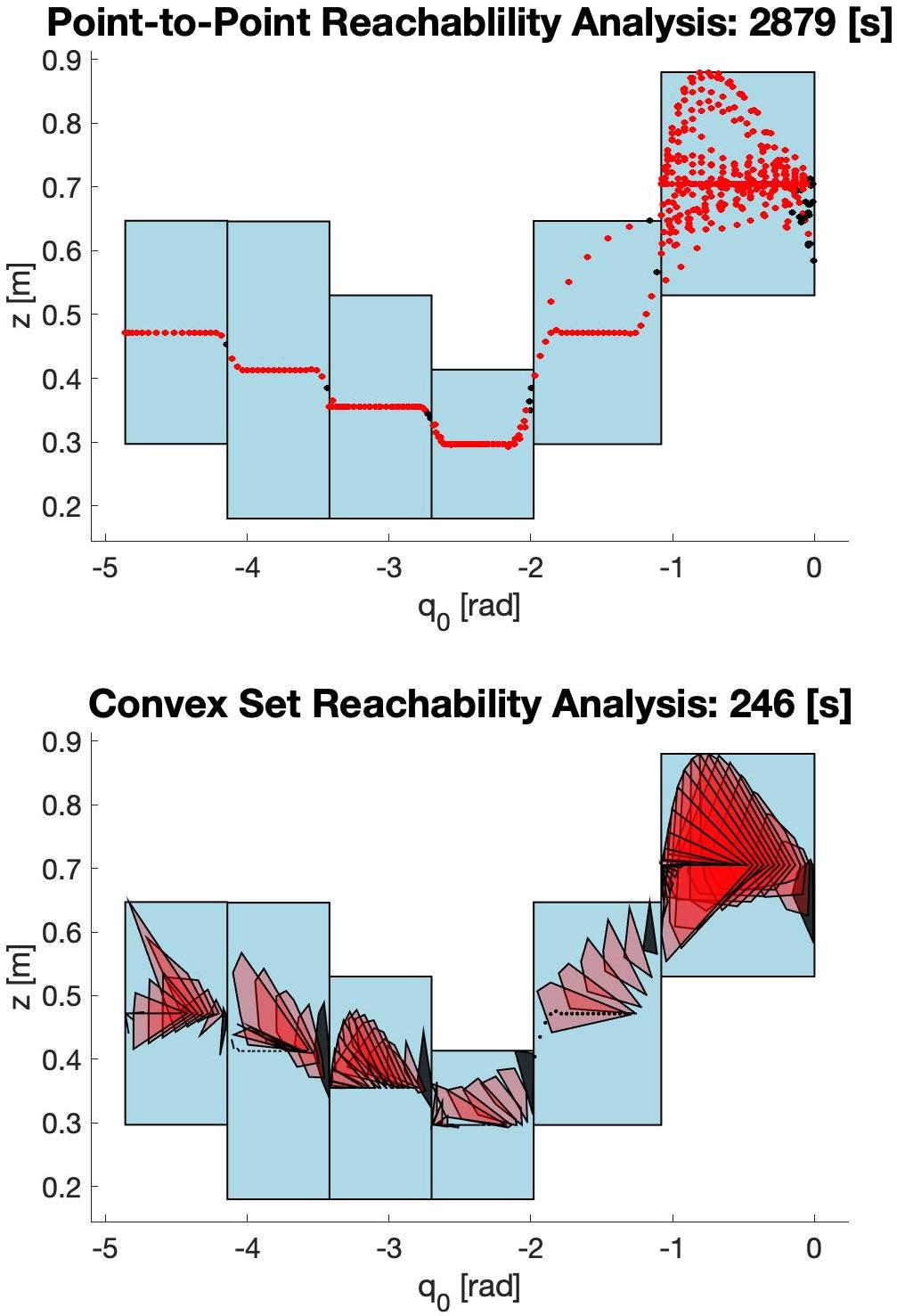}
    \caption{Alg.~\ref{alg:MBTL} produces a significantly larger set of feasible states for $\mathcal{T}2$ in 10\% of the time as the algorithm in \cite{mbttl}. The sampled guard sets for each subtask are plotted in black.}
    \label{fig:robotResults}
\end{figure} 
The new task $\mathcal{T}2$ is configured from another new reshuffling of the obstacles. 
Fig.~\ref{fig:robotResults} depicts the $\mathcal{T}2$ workspace in light blue, along with the $\mathcal{T}2$ safe sets returned by the two versions of the TDMPC algorithm. 
The top image plots in red the feasible safe states for $\mathcal{T}2$ output from the point-to-point controllability method from \cite{mbttl}. 
The bottom image shows in red the feasible safe sets output by the efficient point-to-convex-set controllability method from Alg.~\ref{alg:MBTL}. 

In the example shown, our improved method was an order of magnitude faster at finding safe states for $\mathcal{T}2$ than the point-to-point method, requiring $246$ seconds of processing instead of $2879$ seconds, using a 2017 Mac Book Pro with 2.8 GHz Quad-Core Intel Core i7. 
Indeed, the efficient reformulation of Alg.~\ref{alg:MBTL} for linear systems resulted in an average eleven fold speed-up for five different trials of the described setup and testing procedure. In Tab.~\ref{tb:margins}, each trial corresponds to a newly shuffled $\mathcal{T}2$.

\captionsetup{width=10cm}
\begin{table}[h]
\begin{center}
\caption{Controllability Analysis Run-Time }\label{tb:margins}
\begin{tabular}{ccc}
Trial & Convex Set Analysis & Pointwise Analysis \\\hline \\
1 & 246 s & 2879 s \\
2 & 312 s & 5561 s\\
3 & 219 s& 1618 s\\
4 & 212 s& 1810 s\\
5 & 264 s& 2806 s\\ \hline
\end{tabular}
\end{center}
\end{table}

For each state in a subtask's sampled guard set, the point-to-point controllability method solves a mixed-integer program to try to find an input that controls the system to the last state of a subsequent subtask trajectory. 
Therefore the complexity of both controllability methods depend on the state dimension and number of trajectories through the subsequent subtask (as this provides an upper bound for the size of the subtask safe set).
The efficiency improvement results from replacing the mixed-integer constraint in point-to-point controllability with a convex constraint of equal complexity, which is typically easier to compute. 

\begin{rem}
The convex reachability safe sets in Fig.~\ref{fig:robotResults} demonstrate the implications of Prop.~\ref{prop:1}. Within each subtask, the time-indexed convex safe sets do not entirely contain the previous time step's convex safe set. 
\end{rem}

\begin{rem}
As is clear from Fig.~\ref{fig:robotResults}, the convex set controllability analysis outputs a significantly larger set of feasible states for $\mathcal{T}2$ than the pointwise method. 
This results from two main phenomena.
First, more states from the $\mathcal{T}1$ sampled guard sets remain in $\mathcal{T}2$ when using the convex set controllability than point-to-point controllability. 
This follows since the new controllability analysis (\ref{eq:controllability}) considers a larger one-step target set than the pointwise controllablity analysis (i.e. a convex hull of states in the next subtask rather than individual states). So more points in the sampled guard set can be shown to lead to feasible executions of $\mathcal{T}2$.
Second, while the point-to-point controllability analysis only checks for $\mathcal{T}2$ feasibility of the actual subtask trajectories from $\mathcal{T}1$, the new convex set controllability analysis automatically also provides a policy for the convex subtask safe sets induced by the trajectories. 
All safe states found using the point-to-point controllability method are thus also found using the convex set controllability method.
Accordingly, an ILMPC (\ref{eq:minTimeLMPC} - \ref{eq:ilmpc-policy}) initialized with safe sets returned from Alg.~\ref{alg:MBTL} will also lead to a faster first execution of $\mathcal{T}2$.  
\end{rem}

\section{Conclusion}
In this paper, an extension to the Task Decomposition for iterative learning Model Predictive Control (TDMPC) is presented. 
The algorithm uses stored state and input trajectories from executions of a task, and efficiently designs set-based policies for executing variations of that task. 
The algorithm is designed for linear time-varying systems with piecewise convex constraint sets, and is shown to significantly reduce the computational burden associated with the TDMPC algorithm.
We prove that the resulting policies are guaranteed to be feasible for the new task.
Finally, we evaluate the effectiveness of the proposed algorithm on a robotic path planning tasks, and demonstrate the reduced computational burden compared with TDMPC for nonlinear systems. 

\bibliography{main}             

\appendix
\section{Proof of Proposition~\ref{prop:1}}    
We know that all states in a feasible subtask execution (\ref{eq:subtaskexec}) are controllable to a point in the subtask transition set (\ref{eq:subtasktrans}). 
A sufficient condition for all states in the convex hull of feasible subtask executions to also be controllable to the convex hull of corresponding points in the subtask transition set is for latter to be a control invariant set.

By definition of a feasible task execution (\ref{eq:taskexecution}) and the proof of  Thm.~\ref{th1}, any point in the convex hull of stored states in the transition set is also controllable to the task's control invariant goal set. 
For linear time-invariant systems, states that are controllable to an invariant set are also invariant (\cite{borrelliText}). 
However, for linear time-varying systems these states are only stabilizable to the invariant.
Therefore the convex hull of the points in the transition set is not inherently an invariant.

\begin{figure}[ht]
    \centering
    \includegraphics[width=\columnwidth]{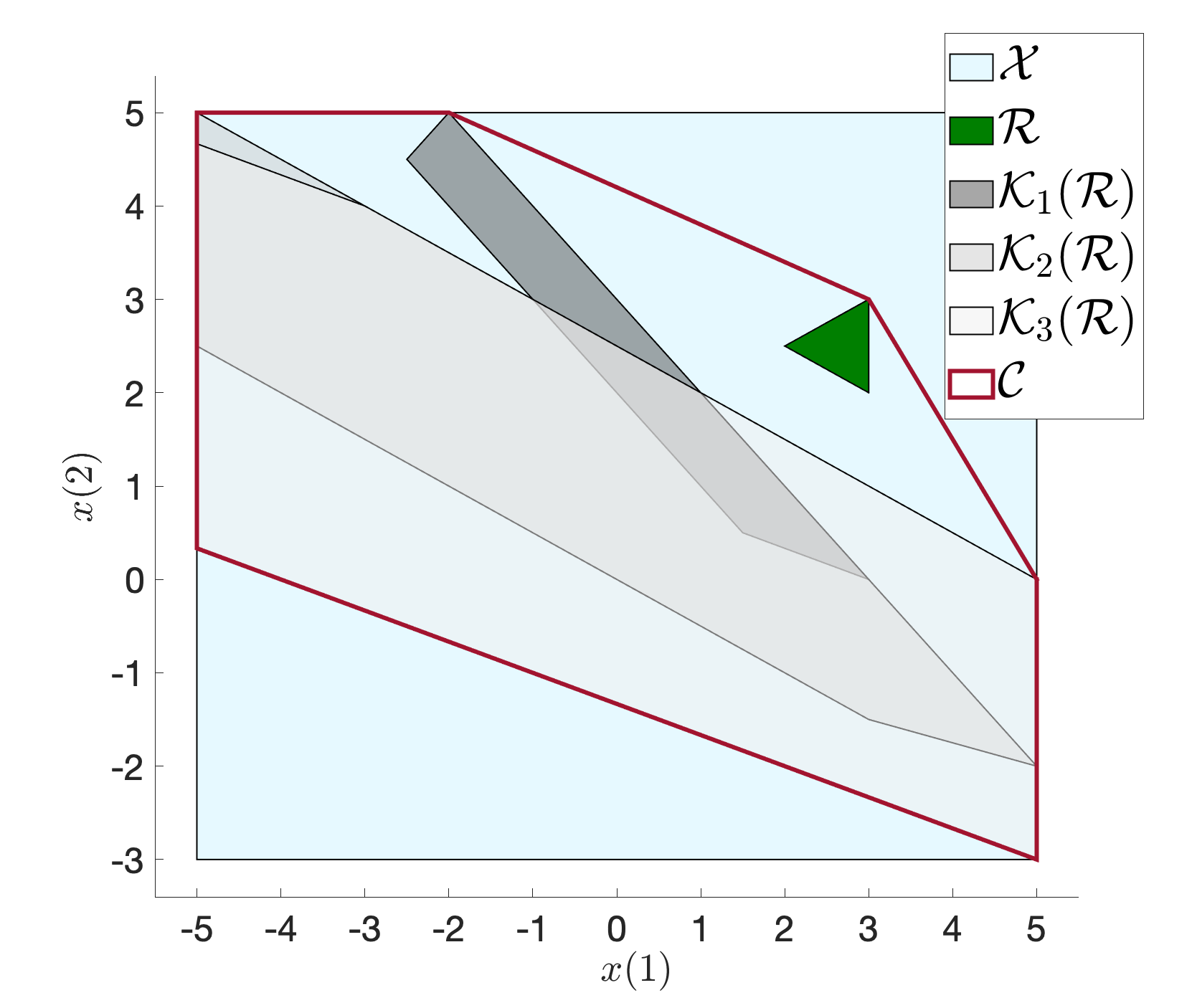}
    \caption{For the double integrator system (\ref{eq:appendixsystem} - \ref{eq:appendixconstr}), the chosen target set (\ref{eq:appendixtarget}) is not an invariant set, as the $N$-step controllable sets are not subsets of the $(N+1)$-step controllable sets.}
    \label{fig:my_label}
\end{figure}

We recognize that the convex hull of states in the transition set \textit{is} a control invariant set if its backward controllable sets grow to contain each other, i.e. if for all scalar values $N$, $\mathcal{K}_{N-1} \subseteq \mathcal{K}_N$.
Unfortunately, this property does not generally hold for linear systems.
Consider for example the double integrator system
\begin{align}\label{eq:appendixsystem}
    x_{k+1} = \begin{bmatrix} 1 & 0 \\ 0 & 1 \end{bmatrix}x_k  + \begin{bmatrix} 0 \\ 1 \end{bmatrix}u_k,
\end{align}
subject to the state and input constraints
\begin{align}\label{eq:appendixconstr}
    x_k \in \mathcal{X}, ~~
    u_k \in [-2, 2].
\end{align}
Define a target set, $\mathcal{R}$, to be the convex hull of three points:
\begin{align}\label{eq:appendixtarget}
    \mathcal{R} = \mathrm{convhull}\bigg( \begin{bmatrix}3 \\ 2 \end{bmatrix} ,  \begin{bmatrix}2 \\ 2.5 \end{bmatrix}, \begin{bmatrix}3 \\ 3 \end{bmatrix}\bigg).
\end{align}
The $1$-step, $2$-step, and $3$-step controllable sets to $\mathcal{R}$ are plotted in Fig.~\ref{fig:my_label}. We also plot $\mathcal{C}$, the convex hull of $\mathcal{R}$ and the controllable sets.
It is clear from the plot that $\mathcal{R}$ is not an invariant set. 
This means it is possible that states in the convex hull of trajectories leading into $\mathcal{R}$ are not $N$-step controllable to $\mathcal{R}$ for any value of $N$. 
In Fig.~\ref{fig:my_label}, this corresponds to states who are in $\mathcal{C}$, but not in any $\mathcal{K}_N(\mathcal{R})$.

Thus, we have shown that the convex hull of stored subtask executions is not generally controllable to the subtask transition set. \hfill$\blacksquare$

\end{document}